\documentclass[11pt]{article}
\usepackage{graphicx}
\usepackage{amsfonts,amsbsy}
\def\Gammabol{{\stackrel{\circ}{\Gamma}}{}}

\def\Rbol{{\stackrel{\circ}{R}}{}}

\def\Gammabol{{\stackrel{\circ}{\Gamma}}{}}

\def\nabol{{\stackrel{\circ}{\nabla}}{}}
\def\Lbol{{\stackrel{\circ}{\mathcal L}}{}}
\def\Gammaw{{\stackrel{\bullet}{\Gamma}}{}}

\def\jw{{\stackrel{~\bullet}{j}}{}}
\def\tw{{\stackrel{\bullet}{t}}{}}

\def\Lw{{\stackrel{\bullet}{\mathcal L}}{}}
\def\Tw{{\stackrel{\bullet}{T}}{}}
\def\Kw{{\stackrel{\bullet}{K}}{}}
\def\nablaw{{\stackrel{\bullet}{\nabla}}{}}

\def\dw{{\stackrel{\bullet}{D}}{}}
\def\Sw{{\stackrel{\bullet}{S}}{}}

\def\onehalf{{\textstyle{\frac{1}{2}}}}

\begin{document}

\renewcommand{\thefootnote}{\fnsymbol{footnote}}
\noindent
{\Large \bf Gravity and the Quantum: Are they Reconcilable?}
\vskip 0.7cm
\noindent
{\bf R. Aldrovandi, J. G. Pereira and K. H. Vu}
\vskip 0.2cm \noindent
{\it Instituto de F\'{\i}sica Te\'orica},
{\it Universidade Estadual Paulista} \\
{\it Rua Pamplona 145},
{\it 01405-900 S\~ao Paulo, Brazil}
%\email{ra@ift.unesp.br}
%\email{jpereira@ift.unesp.br}
%\email{vu@ift.unesp.br}
\vskip 0.8cm
\begin{abstract}
\noindent
General relativity and quantum mechanics are conflicting theories. The seeds of discord are
the fundamental principles on which these theories are grounded. General relativity, on one
hand, is based on the equivalence principle, whose strong version  establishes the {\it
local} equivalence between gravitation and inertia. Quantum mechanics, on the other hand,
is fundamentally based on the uncertainty principle, which is essentially {\it nonlocal} in
the sense that a particle does not follow one trajectory, but infinitely many trajectories,
each one with a different probability. This difference precludes the existence of a quantum
version of the {\it strong} equivalence principle, and consequently of a quantum version of
general relativity. Furthermore, there are compelling experimental evidences that a quantum
object in the presence of a gravitational field violates the {\it weak} equivalence
principle. Now it so happens that, in addition to general relativity, gravitation has an
alternative, though equivalent description, given by teleparallel gravity, a gauge theory
for the translation group. In this theory torsion, instead of curvature, is assumed to
represent the gravitational field. These two descriptions lead to the same classical
results, but are conceptually different. In general relativity, curvature {\it geometrizes}
the interaction, while torsion in teleparallel gravity acts as a {\it force}, similar to
the Lorentz force of electrodynamics. Because of this peculiar property, teleparallel
gravity describes the gravitational interaction without requiring any of the equivalence
principles. The replacement of general relativity by teleparallel gravity may, in
consequence, lead to a conceptual reconciliation of gravitation with quantum mechanics.
\end{abstract}

%%%%%%%%%%%%%%%%%%%%%%
\section{Introduction}
%%%%%%%%%%%%%%%%%%%%%%
%\setcounter{footnote}{0}
%\renewcommand{\thefootnote}{\arabic{footnote}}

%%%%%%%%%%%%%%%%%%%%%%%
\subsection{General Relativity and Universality}
%%%%%%%%%%%%%%%%%%%%%%%

At least at the classical level, gravitation shows a quite peculiar property: 
particles with different masses and different compositions feel it in such a way that all
of them acquire the same acceleration and, given the same initial conditions, follow the 
same path. Such universality of response---usually referred to as {\it universality of
free  fall}---is the most fundamental characteristic of the gravitational interaction
\cite{mtw}. It is unique, peculiar to gravitation: no other basic interaction
of Nature has  it. Effects equally felt by all bodies were known since long. They are the
so called {\it inertial} effects, which show up in non-inertial frames. Examples on Earth
are the centrifugal and the Coriolis forces.

Universality of both gravitational and inertial effects was one of the clues used by
Einstein in building up  general relativity, his theory for gravitation. Another ingredient
was the notion of  field. That concept provides the best approach to interactions coherent
with special relativity.  All known forces are mediated by fields on spacetime. If
gravitation is to be represented by  a field, it should, by the considerations above, be a
universal field, equally felt by  every particle. A natural solution is to assume that
gravitation changes spacetime itself.  And, of all the fields present in a spacetime, the
metric appears as the most fundamental. The  simplest way to change spacetime, then, would
be to change its metric. Furthermore, the  metric does change when looked at from a
non-inertial frame, in which case the (also universal) inertial effects are present. The
presence of a gravitational field should be, therefore, represented by a change in the
spacetime metric. In absence of gravitation that metric should reduce to the flat Minkowski
metric.

A crucial point of Einstein's description, which is fundamentally based on the universality
of free fall, is that it makes no use of the concept of {\em force} for the gravitational
interaction. In fact, instead of acting through a force, gravitation is represented by a
deformation of the spacetime structure. More  precisely, the presence of a gravitational
field is supposed to produce a {\em curvature} in  spacetime, a (spinless) particle in a
gravitational field simply follows a geodesics of the modified spacetime. Notice that no
other kind of spacetime  deformation is supposed to exist. Torsion, for example, which
would be another natural spacetime deformation, is assumed to vanish from the very
beginning. This is the approach of general relativity, in which geometry replaces the
concept of gravitational force, and the trajectories are determined, not by force
equations, but by geodesics. The underlying spacetimes are pseudo-Riemannian spaces.

It is important to remark that only an interaction presenting the property of universality
can be described  by such a geometrization of spacetime. In the eventual absence of 
universality, the general relativity description of gravitation would break down. It is
also important to observe that universality of free fall is usually identified as the
statement of the weak equivalence principle. In fact, if all particles move along
geodesics, the motion will be independent of their masses, and consequently universal. But,
in order to be independent of the masses, they must be somehow canceled out from the
equation of motion. Since this cancellation can only be made when the inertial and
gravitational masses coincide, this last statement is also usually identified with the weak
equivalence principle. It should be remarked, however, that this is true only at the
classical level. At the quantum level, as we are going to see, even if the inertial and
gravitational masses coincide, the gravitational effects on quantum objects can still be
mass-dependent.

%%%%%%%%%%%%%%%%%%%%%%%%%%%%%%%
\subsection{Equivalence Versus Uncertainty Principles}
%%%%%%%%%%%%%%%%%%%%%%%%%%%%%%%

General relativity and quantum mechanics are not consistent with each other. This conflict
stems from the very principles on which these theories take their roots. General relativity,
on one hand, is based on the equivalence principle, whose strong version establishes the
{\it local} equivalence between gravitation and inertia. The fundamental asset of quantum
mechanics, on the other hand, is the uncertainty principle, which is essentially {\it
nonlocal}: a test particle does not follow a given trajectory, but infinitely many
trajectories, each one with a different probability \cite{chiao}. A crucial question then
arises: is there a peaceful way of reconciling the equivalence and the uncertainty
principles? The answer seems to be {\it no} as these two principles  are fundamentally
different, and like darkness and lightness, they cannot hold  simultaneously. It then comes
the inevitable question: which one is to be discarded? At first sight the answer seems to
be very difficult because general relativity and quantum mechanics are two of the main
pillars of modern physics, and discarding one of their underlying principles would mean to
discard one of these pillars. However, a more careful analysis of this question strongly
suggests that the equivalence principle is the weaker part of the building.

To begin with, we note that the strong version of the equivalence principle, which requires
the weak one, presupposes an ideal observer \cite{ABP02}, represented by a timelike curve
which intersects the space-section {\em at a point}. In each space-section, it applies at
that intersecting point. The conflict comes, for the strong principle, from that
idealization and extends, clearly, also to special relativity. In the equation for a curve,
gravitation only appears through the Levi--Civita connection, which can be made to vanish
all along. An ideal observer can choose frames whose acceleration exactly compensate the
effect of gravitation. A real observer, on the other hand, will be necessarily an object
extended in space, consequently intersecting a congruence of curves. Such congruences are
described by the deviation equation and, consequently, detect the true covariant object
characterizing the gravitational field, the curvature tensor which cannot be made to
vanish. Quantum Mechanics requires real observers, pencils of ideal observers. The
inconsistency with the strong principle, therefore, is a mathematical necessity. It is not
possible, as a consequence, to define a quantum version of the strong equivalence
principle \cite{qsep}. On the other hand, the inconsistency of quantum mechanics with the
weak equivalence principle is a matter of experiment. Although it has passed all
experimental tests at the classical level \cite{exp}, as we are going to see, there are
compelling evidences that the weak equivalence principle might not be true at the quantum
level.

It should be mentioned that, even at the classical level, there are many controversies
related with the correct meaning of the strong equivalence principle. For example, in the
Preface of his classic textbook \cite{synge}, Synge confess that ... {\it I have never been
able to understand this Principle. Does it mean that the signature of the space-time metric
is $+2$ (or $-2$ if you prefer the other convention)? If so, it is important, but hardly a
Principle. Does it mean that the effects of a gravitational field are indistinguishable
from the effects of an observer's acceleration? If so, it is false. In Einstein's theory,
either there is a gravitational field or there is none, according to as the Riemann tensor
does not or does vanish. This is an absolute property; it has nothing to do with any
observer's world line. Space-time is either flat or curved, and in several places in the
book I have been at considerable pains to separate truly gravitational effects due to
curvature of space-time from those due to curvature of the observer's world-line (in most
ordinary cases the latter predominate). The Principle of Equivalence performed the
essential office of midwife at the birth of general relativity, but, as Einstein remarked,
the infant would never have got beyond its long-clothes had it not been for Minkowski's
concept. I suggest that the midwife be now buried with appropriate honours and the facts of
absolute space-time faced}. Many other criticisms can be found in the literature
\cite{damour,bondi}, but will not consider them here as our main interest will be focused on
what happens at the quantum level.

%%%%%%%%%%%%%%%%%%%%%%%%%%%%%%%
\subsection{Purposes}
%%%%%%%%%%%%%%%%%%%%%%%%%%%%%%%

Following the arguments described above, the basic purpose of this paper will be to explore
further the conceptual conflict between general relativity and quantum mechanics. Then,
by considering the teleparallel equivalent of general relativity, which has already been
shown not to require any of the equivalence principles to describe the gravitational
interaction \cite{wep}, a possible way of solving the conflict, and consequently
reconciling gravitation with quantum mechanics, will be proposed and analyzed. We begin by
presenting, in the next section, some evidences that the weak equivalence principle fails
at the quantum level.

%%%%%%%%%%%%%%%%%%%%
\section{Quantum Effects in Gravitation}
%%%%%%%%%%%%%%%%%%%%

\begin{figure}[htbp]
\begin{center}
\includegraphics[width=7.2cm]{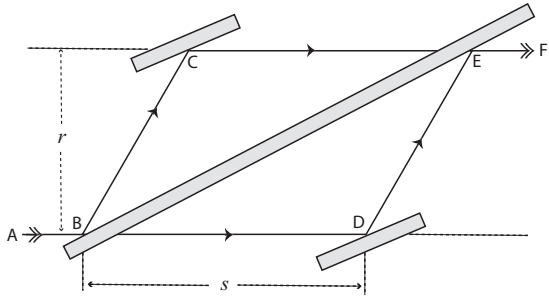}
\caption{Schematic illustration of the COW neutron interferometer.}
\end{center}
\label{fig1}
\end{figure}

There are in the literature some very simple idealized examples of possible quantum
violations of the weak equivalence principle \cite{viola}. Here, however, we will present
two very specific evidences, the first of them extensively verified experimentally. It is
the so called Colella--Overhauser--Werner (COW) phenomenon \cite{cow}. It consists in using
a neutron interferometer to observe the quantum mechanical phase shift of neutrons caused
by their interaction with Earth's gravitational field, assumed to be Newtonian.
Figure~\ref{fig1} shows a scheme of the experiment, which is performed in the presence of a
Newtonian potential
\begin{equation}
\phi \equiv g \, z,
\end{equation}
where $g$ is the gravitational acceleration, supposed not to change significantly in the
region of the experience, and $z$ is the distance from some reference point on Earth.
In the presence of a gravitational field, because the segments {\sf BD} and {\sf  CE} are
at different distance from Earth, and consequently at different value of the  potential
$\phi$, there will be a gravitationally induced quantum phase shift between the  two
trajectories when they arrive at the screen. This phase shift is given by \cite{cow}
\begin{equation}
\delta \varphi \equiv \varphi_{\sf BCE} - \varphi_{\sf BDE} =
\frac{g r s}{\hbar v} \; m,
\end{equation}
where $r$ and $s$ are dimensions of the interferometer (see Fig.~\ref{fig1}), $v$ is the
velocity, and $m$ is the mass of the neutron. From this expression we can see that the
quantum phase difference induced by the gravitational field depends explicitly on the
mass of the particle. More specifically, if we distinct the gravitational $(m_g$) and
inertial $(m_i)$ masses, the phase difference in this case would be
\cite{global,caxambu}
\begin{equation}
\delta \varphi =
\frac{g r s}{\hbar v} \; m_g,
\end{equation}
from where we see that, actually, the phase shift depends on the gravitational mass of the
particle. At the quantum level, therefore, due to this dependence, gravitation seems to be
no more universal \cite{nonuni}.\footnote{It should be remarked that, through the
definition of a quantum version of the weak equivalence principle \cite{quantu}, the
phase shift of non-relativistic interferometry experiments can be made independent of the
mass if written in an appropriate way. It is not clear, however, whether this principle
remains valid in the relativistic domain.}

%%%%%%%%%%%%%%%%%%%
%%%%%%%%%%%%%%%%%%%

As another evidence of a possible quantum violation of universality, let us consider now
the gravitational analog of the Aharonov--Bohm effect \cite{gab}. The usual
(electromagnetic) Aharonov--Bohm effect consists in a shift, by a constant amount, of the
electron interferometry  wave pattern, in a region where there is no magnetic field, but
there is a nontrivial electromagnetic potential. Analogously, the gravitational
Aharonov--Bohm effect  will consist in a similar shift of the same wave pattern, but
produced by the presence of a  gravitational potential, in a region where there is no
gravitational field. Phenomenologically,  this kind of effect might be present near a
massive rapidly rotating source, like a neutron  star, for example. 
Of course, differently from an ideal apparatus, in a real situation the gravitational
field cannot be completely eliminated, and consequently the  gravitational Aharonov--Bohm
effect should be added to the other effects also causing a phase change.

%%%%%%%%%%%%%%%%%%%
\begin{figure}[htbp]
\begin{center}
\includegraphics[width=7.0cm]{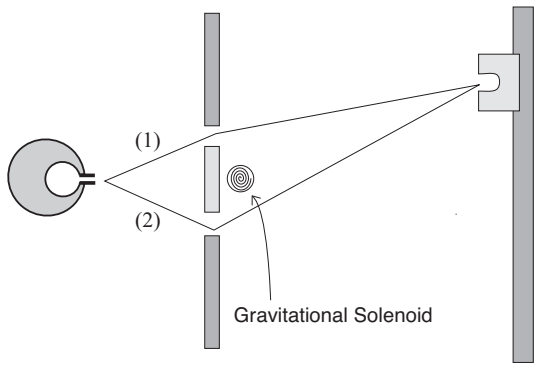}
\caption{Schematic illustration of the Aharonov--Bohm electron interferometer.}
\end{center}
\label{fig2}
\end{figure}
%%%%%%%%%%%%%%%%%%%

We consider then the ideal case in which a kind of infinite ``gravitational solenoid''
produces a purely static gravitomagnetic field flux concentrated in its interior (see
Fig.~\ref{fig2}). In the ideal situation, the gravitational field outside the solenoid
vanishes completely, but there is a nontrivial gravitational potential. When we  let the
electrons to move outside the solenoid, phase factors corresponding to paths lying on one
side of the solenoid will interfere with phase factors corresponding to paths lying on
the other side, which will produce an additional phase shift at the screen. Denoting by
\begin{equation}
\Omega = \oint \vec{H} \cdot d\vec{\sigma}
\end{equation}
the flux of the gravitomagnetic field $\vec{H}$ inside the solenoid, the phase  shift is
found to be \cite{global}
\begin{equation}
\delta \varphi \equiv \varphi_{(2)} - \varphi_{(1)} = \frac{{\mathcal E} \, 
\Omega}{\hbar \,
c},
\label{gabe2}
\end{equation}
where ${\mathcal E} = \gamma m c^2$ is the electron kinetic energy, with $\gamma \equiv
[1 - ({v^2}/{c^2}) ]^{-1/2}$ the relativistic factor. As it depends on the energy, this
phase difference applies equally to massive and massless  particles. For the case of
massive particles, if we distinct gravitational and inertial masses, the phase shift
would be \cite{caxambu}
\begin{equation}
\delta \varphi = \frac{{\mathcal E} \, \Omega}{\hbar \, c} \left(\frac{m_g}{m_i} \right)
= \frac{\gamma  c \, \Omega}{\hbar} \; m_g,
\label{gabe2bis}
\end{equation}
where now ${\mathcal E} = \gamma m_i c^2$. We see from this expression that, also in the
gravitational analog of the Aharonov--Bohm effect, the phase shift depends on the
(gravitational) mass of the particle. This is one more indication that, at the quantum
level, gravitation seems to be no more universal.

%%%%%%%%%%%%%%%
\section{Teleparallel Gravity}
%%%%%%%%%%%%%%%

%%%%%%%%%%%%%%%%%%%%%%%%%%%%%
\subsection{Can We Dispense with the Weak Equivalence Principle?}
%%%%%%%%%%%%%%%%%%%%%%%%%%%%%

The basic conclusion of the previous section was that there are strong indications that
gravitation is no more universal at the quantum level. This means essentially that the
weak equivalence principle is no more applicable at this level. However, as already
discussed, without this principle, the geometrical description of general relativity
breaks down. A new question then arises: are we able to manage without the equivalence
principle, and consequently without general relativity? The remaining of this paper will
be devoted to answer this question.

To begin with, let us remark that, like the other fundamental interactions of nature,
gravitation can also be described in terms of a gauge theory \cite{hehl}. In fact, the 
teleparallel equivalent of general relativity, or teleparallel gravity for
short,\footnote{The name teleparallel gravity is normally used to designate a theory in
which there are three free parameters (see, for example, Ref.~\cite{op1}, and references
therein). Here, however, we use it as a synonymous of the teleparallel equivalent of
general relativity, a theory obtained for a specific choice of these parameters.} can be
interpreted as a gauge theory for the translation group. In this theory, instead of
torsion, curvature is  assumed to vanish. The corresponding underlying spacetime is, in
this case, a Weitzenb\"ock spacetime \cite{weitz}. In spite of this fundamental
difference, the two theories are found to yield equivalent classical descriptions of the
gravitational interaction \cite{equiva}. Conceptual differences, however, show up.
According to general relativity, curvature is used to {\it geometrize} spacetime.
Teleparallelism, on the other hand, attributes gravitation to torsion, but in this case
torsion accounts for gravitation not by geometrizing the interaction, but by acting as a
{\it force}. As a  consequence, there are no geodesics in teleparallel gravity, but only
force equations quite analogous to the Lorentz force equation of electrodynamics
\cite{paper1}. We may then say that the gravitational interaction can be described in
terms of curvature, as is usually done in general relativity, or {\em alternatively} in
terms of torsion, in which case we  have the so called teleparallel gravity. Whether
gravitation requires a curved or a torsioned connection---or equivalently, a Riemann or a
Weitzenb\"ock spacetime structure---turns out to be, at least classically, a matter of
convention.

One may wonder why a gauge theory for the translation group, and not for other spacetime
group. The reason for this is related to the source of gravitation, that is, energy and
momentum. As is well known from Noether's theorem \cite{kopo}, these quantities are
conserved provided the Lagrangians are invariant under spacetime translations. It is
then natural to expect that the gravitational field be represented by a gauge theory for
the translation group. This is quite similar to the electromagnetic field, whose
source---the electric four-current---is conserved due to invariance of the source Lagrangian
under transformations of the unitary group U(1), the gauge group of Maxwell's theory. Observe
that angular momentum itself (either orbital or intrinsic) is not source of gravitation:
only the energy and momentum associated to that angular momentum can be source of
gravitation. This means that Lorentz transformations cannot play any dynamical role in a
gauge approach to gravitation. The role played by the Lorentz transformations is the same
they play in special relativity: the {\it local} Lorentz group provides a relation between
different classes of frames, each class defined by all frames equivalent under {\it global}
Lorentz transformations. Physics, of course, cannot depend on the frame used to describe it.
Not only gravitation, therefore, but all theories must be invariant under local Lorentz
transformations. 

Now, as is widely known, the electromagnetic interaction is not universal: there  exists no
electromagnetic equivalence principle. As both Maxwell's theory and teleparallel gravity
are Abelian gauge theories in which the equations of motion of test particles are not
geodesics but force equations, the question arises whether the gauge approach of
teleparallel gravity would also be able to describe the gravitational interaction in the
eventual lack of universality. As we are going to see, the answer to this question is {\em
yes}: teleparallel gravity does not require the validity of the equivalence principle to
describe the gravitational interaction \cite{wep}. Whereas the geometrical description of
general relativity breaks down in the absence of universality, teleparallel gravity remains
a consistent theory. In spite of the equivalence with general relativity, therefore,
teleparallel gravity seems to belong to a more general class of theory. In order to
understand this point, it is necessary first to study the fundamentals of teleparallel
gravity.

%%%%%%%%%%%%%%%%%%%%%%%%%
\subsection{Fundamentals of Teleparallel Gravity}
%%%%%%%%%%%%%%%%%%%%%%%%%

The mathematical structure of distant parallelism, also referred to as absolute or
teleparallelism, was used by Einstein in the late nineteen twenties, in his  attempt to
unify gravitation with electromagnetism. The crucial idea was the introduction of a 
tetrad field, a field of orthonormal bases of the tangent spaces at each point of the
four-dimensional spacetime. The specification of a tetrad involves sixteen components,
whereas the gravitational field, represented by the spacetime metric, requires only ten 
components. The six additional degrees of freedom ensued by the tetrad was then supposed by 
Einstein to represent the electromagnetic field. This attempt of unification did not succeed,
but some of the concepts introduced by him remain important up to the present day \cite{uni}.

According to the gauge structure of teleparallel gravity, to each point of spacetime there
is attached a Minkowski tangent space, on which the translation (gauge) group acts. We use
the Greek alphabet $\mu, \nu, \rho, \dots = 0, 1, 2, 3$ to denote spacetime indices and
the Latin alphabet $a, b, c, \dots = 0, 1, 2, 3$ to denote algebraic indices related  to
the tangent Minkowski spaces, whose metric is chosen to be $\eta_{a b} = {\rm diag}  (+1,
-1, -1, -1)$. As a gauge theory for translations, the fundamental field of teleparallel 
gravity is the translational gauge potential $B^a{}_\mu$, a 1-form assuming values in the
Lie  algebra of the translation group
\begin{equation}
B_\mu = B^a{}_\mu \, P_a,
\end{equation}
with $P_a = \partial_a$ the generators of infinitesimal translations. Under a local
translation of the tangent space coordinates $\delta x^a = \epsilon^a(x) \equiv 
\epsilon^a$, the gauge potential transforms according to
\begin{equation}
B^{\prime a}{}_\mu = B^a{}_\mu - \partial_\mu \epsilon^a.
\label{btrans}
\end{equation}
It appears naturally as the nontrivial part of the tetrad field $h^{a}{}_{\mu}$:
\begin{equation}
h^a{}_\mu = \partial_\mu x^a + B^a{}_\mu.
\label{tetrada}
\end{equation}
If the tangent space indices are raised and lowered with the Minkowski
metric $\eta_{a b}$, therefore, the spacetime indices will raised and lowered with 
the
spacetime metric
\begin{equation}
g_{\mu \nu} = \eta_{a b} \; h^a{}_\mu \, h^b{}_\nu.
\label{gmn}
\end{equation}
The above tetrad can be used to construct the so called Weit\-zen\-b\"ock connection
\begin{equation}
\Gammaw^{\rho}{}_{\mu\nu} = h_{a}{}^{\rho}\partial_{\nu}h^{a}{}_{\mu},
\label{carco}
\end{equation}
which introduces the distant parallelism in the four-dimensional spacetime  manifold. It
is a connection presenting torsion, but no curvature. Its torsion,
\begin{equation}
\Tw^{\rho}{}_{\mu\nu} = \Gammaw^{\rho}{}_{\nu\mu} - 
\Gammaw^{\rho}{}_{\mu\nu},
\label{tor}
\end{equation}
is nothing but the translational gauge field strength $\Tw^a{}_{\mu \nu}$, as seen from the
tetrad frame:
\begin{equation}
\Tw^a{}_{\mu \nu} \equiv \partial_\mu B^a{}_{\nu} - \partial_\nu B^a{}_{\mu} =
h^a{}_\rho \; \Tw^\rho{}_{\mu \nu}.
\label{gfs}
\end{equation}
The Weitzenb\"ock connection is related to the Levi--Civita connection
\begin{equation}
{\stackrel{\circ}{\Gamma}}{}^{\rho}{}_{\mu\nu} = {\textstyle
\frac{1}{2}} \, g^{\rho \sigma} \left( \partial_{\mu} g_{\sigma \nu} +
\partial_{\nu} g_{\sigma \mu} - \partial_{\sigma} g_{\mu \nu} \right)
\label{lci}
\end{equation}
of the spacetime metric $g_{\mu\nu}$ through
\begin{equation}
\Gammaw^{\rho}{}_{\mu\nu} = \Gammabol^{\rho}{}_{\mu\nu} + \Kw^{\rho}{}_{\mu\nu},
\label{rela}
\end{equation}
where
\begin{equation}
\Kw^{\rho}{}_{\mu \nu} = \textstyle{\frac{1}{2}} \left( 
\Tw_{\mu}{}^{\rho}{}_{\nu} + \Tw_{\nu}{}^{\rho}{}_{\mu} 
- \Tw^{\rho}{}_{\mu \nu} \right)
\label{contorsion}
\end{equation}
is the contortion tensor.

The Lagrangian of the teleparallel equivalent of general relativity is \cite{paper1}
\begin{equation}
\Lw = \frac{h}{8 k^2} \left[
\Tw^\rho{}_{\mu \nu} \Tw_\rho{}^{\mu \nu} + 2 \,
\Tw^\rho{}_{\mu \nu} \Tw^{\nu \mu} {}_\rho - 4 \, \Tw_{\rho \mu}{}^{\rho}
\Tw^{\nu \mu}{}_\nu \right].
\label{lagr3}
\end{equation}
where $k^2 = 8 \pi G/c^{4}$ and $h \equiv \sqrt{-g} = {\rm det}(h^{a}{}_{\mu})$, with
$g = \det(g_{\mu \nu})$. The first term corresponds to the usual Lagrangian of  {\em
internal}, or Yang--Mills gauge theories. In the gravitational case, however, owing to the 
presence of a tetrad field, which are components of the solder form \cite{livro}, algebra
and spacetime indices can be changed into each other, and in consequence new contractions
turn out to be possible. It is exactly this  possibility that gives rise to the other two
terms of the above Lagrangian. Defining the tensor
\begin{equation}
\Sw^{\rho\mu\nu} = - \Sw^{\rho\nu\mu} =
\left[ \Kw^{\mu\nu\rho} - g^{\rho\nu}\,\Tw^{\sigma\mu}{}_{\sigma} 
+ g^{\rho\mu}\,\Tw^{\sigma\nu}{}_{\sigma} \right],
\label{S}
\end{equation}
usually called superpotential \cite{mollersbook}, it can be rewritten in the form
\cite{maluf}
\begin{equation}
\Lw =
\frac{ h}{4 k^2} \; \Tw^\rho{}_{\mu\nu} \, \Sw_\rho{}^{\mu\nu}.
\label{gala}
\end{equation}

Let us consider now the Lagrangian
\begin{equation}
{\mathcal L} = \Lw + {\mathcal L}_m,
\end{equation}
where ${\mathcal L}_m$ represents the Lagrangian of a general matter field. By performing
variations in relation to the gauge field $B^a{}_\rho$, we obtain the teleparallel version
of the gravitational field equation
\begin{equation}
\partial_\sigma(h \Sw_a{}^{\rho \sigma}) -
k^2 \, (h \jw_{a}{}^{\rho}) = k^2 \, (h {\mathcal T}_{a}{}^{\rho}),
\label{tfe1}
\end{equation}
where $\Sw_a{}^{\rho \sigma} = h_{a}{}^{\lambda} \Sw_{\lambda}{}^{\rho \sigma}$, the current
\begin{equation}
\jw_a{}^\rho \equiv - \frac{\partial {\mathcal \Lw}}{\partial h^a{}_\rho} =
\frac{h_a{}^\lambda}{k^2}
\left( \Tw^c{}_{\mu \lambda} \, \Sw_c{}^{\mu \rho} - \frac{1}{4} \, 
\delta_\lambda{}^\rho \,
\Tw^c{}_{\mu \nu} \, \Sw_c{}^{\mu \nu} \right)
\label{emt1bis}
\end{equation}
represents the tensorial form of the gravitational energy-momentum density \cite{prl}, and
\begin{equation}
h {\mathcal T}_{a}{}^{\rho} \equiv - \frac{\delta {\mathcal L}_m}{\delta
B^a{}_{\rho}} \equiv - \frac{\delta {\mathcal L}_m}{\delta h^a{}_{\rho}} = -
\left( \frac{\partial {\mathcal L}_m}{\partial h^a{}_{\rho}} -
\partial_\lambda \frac{\partial {\mathcal L}_m}{\partial_\lambda\partial 
h^a{}_{\rho}}
\right)
\label{memt1}
\end{equation}
is the matter energy-momentum tensor. Due to the anti-symmetry of $\Sw_a{}^{\rho \sigma}$
in the last two indices, the total current is conserved as  a consequence of the field
equation:
\begin{equation}
\partial_\rho \left[ h \left( \jw_a{}^\rho + {\mathcal T}_{a}{}^{\rho} \right) 
\right] = 0.
\label{conser0}
\end{equation}
In a purely spacetime form, the above field equation acquires the form
\begin{equation}
\partial_\sigma(h \Sw_\lambda{}^{\rho \sigma}) -
k^2 \, (h \tw_\lambda{}^\rho) = k^2 \, (h {\mathcal T}_{\lambda}{}^{\rho}),
\label{eqs1}
\end{equation}
where
\begin{equation}
h \, \tw_\lambda{}^\rho = \frac{h}{k^2} \left( \Gammaw^\mu{}_{\nu\lambda} \,
\Sw_{\mu}{}^{\rho \nu} - \frac{1}{4} \, \delta_\lambda{}^\rho \,
\Tw^\theta{}_{\mu\nu} \, \Sw_\theta{}^{\mu\nu} \right)
\label{emt1}
\end{equation}
is the energy-momentum {\it pseudotensor} of the gravitational field, and 
${\mathcal T}_{\lambda}{}^{\rho} = {\mathcal T}_{a}{}^{\rho} \; h^a{}_\lambda$.
It is important to notice that $\tw_{\lambda}{}^{\rho}$ is not
simply the gauge current $\jw_a{}^\rho$ with the algebraic index ``$a$'' changed to the
spacetime index ``$\lambda$''. It incorporates also an extra term coming from the
derivative term of the field equation:
\begin{equation}
\tw_\lambda{}^\rho = h^a{}_\lambda \, \jw_a{}^\rho +
k^{-2} \, \Gammaw^{\mu}{}_{\lambda \nu} \, \Sw_{\mu}{}^{\rho \nu}.
\label{ptem3}
\end{equation}
We see clearly from this equation the origin of the connection-term which transforms the
gauge current $\jw_a{}^\rho$ into the energy-momentum pseudotensor $\tw_\lambda{}^\rho$.

Now, using the relation (\ref{rela}), it is possible to show that
\begin{equation}
\Lw = \Lbol - \partial_\mu \left(2 \, h \, k^{-2} \,
\Tw^{\nu \mu}{}_\nu \right),
\end{equation}
where
\begin{equation}
\Lbol = - \frac{\sqrt{-g}}{2 k^2} \; \Rbol
\end{equation}
represents the Einstein--Hilbert Lagrangian of general relativity, with $\Rbol$ the  scalar
curvature of the Levi--Civita connection $\Gammabol^{\rho}{}_{\mu\nu}$. Up to a 
divergence, therefore, the teleparallel Lagrangian is equivalent to the Einstein--Hilbert 
Lagrangian of general relativity. It is important to observe also that, by using 
Eq.~(\ref{rela}), the left-hand side of the field equation (\ref{eqs1}) can be shown to
satisfy the  relation
\begin{equation}
\partial_\sigma(h \Sw_\lambda{}^{\rho \sigma}) -
k^2 \, (h \tw_{\lambda}{}^{\rho}) =
{h} \left({\stackrel{\circ}{R}}_\lambda{}^{\rho} -
\onehalf \, \delta_\lambda{}^{\rho} \;
{\stackrel{\circ}{R}} \right).
\label{ident}
\end{equation}
This means that, as expected due to the equivalence between the cor\-re\-sponding
Lagrangians, the teleparallel field equation (\ref{tfe1}) is equivalent to  Einstein's
field equation
\begin{equation}
{\stackrel{\circ}{R}}_\lambda{}^{\rho} -
\onehalf \, \delta_\lambda{}^{\rho} \,
{\stackrel{\circ}{R}} = k^2 \, {\mathcal T}_{\lambda}{}^{\rho}.
\end{equation}
We see in this way that, as already remarked, in spite of the conceptual  differences
between teleparallel gravity and general relativity, these theories are found to yield 
equivalent descriptions of gravitation. Although equivalent, however, they describe the 
gravitational interaction through a completely different mechanism. In the next section we
are  going to explore these differences.

%%%%%%%%%%%%%%%%%%%%%%
\section{Force Equation Versus Geodesics}
%%%%%%%%%%%%%%%%%%%%%%

Let us consider, in the context of teleparallel gravity, the motion of a spinless particle
of mass $m$ in a gravitational field $B^{a}{}_{\mu}$.  Analogously to the electromagnetic
case \cite{ll}, the action integral is written in the form
\begin{equation}
{\mathcal S} = \int_{a}^{b} \left[ - m \, c \, d\sigma -
m \, c \, B^{a}{}_{\mu} \, u_{a} \, dx^{\mu} \right],
\label{acaop1}
\end{equation}
where $d\sigma = (\eta_{a b} dx^a dx^b)^{1/2}$ is the Minkowski tangent-space invariant
interval,
\begin{equation}
u^a = h^a{}_\mu \, u^\mu,
\end{equation}
is the anholonomic particle four-velocity, with
\begin{equation}
u^\mu = \frac{d x^\mu}{ds}
\label{ust}
\end{equation}
the holonomic four-velocity, which is written in terms of the spacetime invariant interval
$ds$ = $(g_{\mu \nu} dx^\mu dx^\nu)^{1/2}$.

The first term of the action (\ref{acaop1}) represents the action of a free particle, and
the second the coupling of the particle's mass with the gravitational field.  Notice that
the separation of the action in these two terms is possible only in a gauge theory, like
teleparallel gravity, being not possible in general relativity. It is, however, equivalent
with the usual action of general relativity. In fact, if we introduce the identities
\cite{wep}
\begin{equation}
h^a{}_\mu u_a u^\mu = 1
\end{equation}
and
\begin{equation}
h^a{}_\mu \; \frac{d \sigma}{d s} = \frac{\partial x^a}{\partial x^\mu},
\end{equation}
the action (\ref{acaop1}) can easily be seen to reduce to its general relativity version
\begin{equation}
{\mathcal S} = - \int_{a}^{b} m \, c \, ds.
\label{graction}
\end{equation}
In this case, the interaction of the particle with the gravitational field is described by
the metric tensor $g_{\mu \nu}$, which is present in $ds$.

Variation of the action (\ref{acaop1}) yields the equation of motion
\begin{equation}
h^a{}_\mu \, \frac{d u_a}{d s} =
\Tw^a{}_{\mu \rho} \; u_a \, u^\rho.
\label{eqmot2}
\end{equation}
This is the force equation governing the motion of the particle, in which the teleparallel
field strength $\Tw^a{}_{\mu \rho}$---that is, torsion---plays the role of gravitational
force. To write it in a purely spacetime form, we use the relation
\begin{equation}
h^{a}{}_{\mu} \frac{d u_{a}}{d s} = \omega_\mu \equiv
\frac{d u_{\mu}}{d s} - \Gammaw^\theta{}_{\mu \nu} u_{\theta} \, u^{\nu},
\end{equation}
where $\omega_\mu$ is the spacetime particle four-acceleration. We then get
\begin{equation}
u^\nu \nablaw_\nu u_\mu \equiv \frac{d u_\mu}{d s} - \Gammaw^\theta{}_{\mu \nu} \;
u_\theta \; u^\nu = \Tw^\theta{}_{\mu \nu} \; u_\theta \; u^\nu.
\label{geode}
\end{equation}
The left-hand side of this equation is the Weitzenb\"ock covariant derivative of
$u_\mu$ along the world-line of the particle. The presence of the torsion tensor on its
right-hand side, as already stressed, shows that in teleparallel gravity torsion plays
the role of gravitational force. By using the identity
\begin{equation}
\Tw^\theta{}_{\mu \nu} \, u_\theta \, u^\nu = - \Kw^\theta{}_{\mu \nu}
\, u_\theta \, u^\nu,
\label{tuukuu}
\end{equation}
this equation can be rewritten in the form
\begin{equation}
u^\nu \dw_\nu u_\mu \equiv \frac{d u_\mu}{d s} - \left(\Gammaw^\theta{}_{\mu \nu} 
- \Kw^\theta{}_{\mu \nu} \right) u_\theta \; u^\nu = 0.
\label{geode3}
\end{equation}
The left-hand side of this equation is the teleparallel covariant derivative of
$u_\mu$ along the world-line of the particle. Using the relation (\ref{rela}), it  is
found to be
\begin{equation}
u^\nu \nabol_\nu u_\mu \equiv \frac{d u_\mu}{d s} -
{\stackrel{\circ}{\Gamma}}{}^\theta{}_{\mu \nu} \; u_\theta \; u^\nu = 0.
\label{geo2}
\end{equation}
This is precisely the geodesic equation of general relativity, which means that the
trajectories followed by spinless particles are geodesics of the underlying  Riemann
spacetime. In a locally inertial coordinate system, the first derivative of the metric
tensor vanishes, the Levi--Civita connection vanishes as well, and the geodesic equation
(\ref{geo2}) becomes the equation of motion of a free particle. This is the usual version
of the (strong) equivalence principle as formulated in general relativity \cite{weinberg}.

It is important to notice that the same principle holds in teleparallel gravity, but it
operates differently. To see that, we use the torsion definition (\ref{tor}) to rewrite the
force equation (\ref{geode}) in the form
\begin{equation}
\frac{d u_\mu}{d s} - \Gammaw^{\theta}{}_{\nu \mu} \; u_\theta \; u^\nu = 0.
\label{geodeflat}
\end{equation}
Observe that, as $\Gammaw_{\theta \nu \mu}$ is not symmetric in the last two indices, the
left-hand side is not the covariant derivative of the four-velocity along the trajectory,
and consequently it is not a geodesic equation. In other words, the trajectories followed by
spinless particles are not geodesics of the underlying Weitzenb\"ock spacetime. In a locally
inertial coordinate system, the first derivative of the metric tensor vanishes, and the
Weitzenb\"ock connection $\Gammaw_{\theta \nu \mu}$ becomes skew-symmetric in the first two
indices. In this coordinate system, therefore, owing to the symmetry of $u^\theta \; u^\nu$,
the force equation (\ref{geodeflat}) becomes the equation of motion of a free particle. This
is the teleparallel version of the (strong) equivalence principle \cite{paper1}.

%%%%%%%%%%%%%%%%%%%%%%%%%%%%%%
\section{Managing without the Equivalence Principle}
%%%%%%%%%%%%%%%%%%%%%%%%%%%%%%

As an example of an explicit violation of the weak equivalence principle, let us consider
again the motion of a spinless particle in a gravitational field represented by the
translational gauge potential $B^{a}{}_{\mu}$, but assuming that the gravitational ($m_g$)
and inertial ($m_i$) masses do not coincide. In this case, the teleparallel version of the
action is written in the form
\begin{equation}
{\mathcal S} = \int_{a}^{b} \left( - m_i \, c \, d\sigma -
m_g \, c \, B^{a}{}_{\mu} \, u_{a} \, dx^{\mu} \right).
\label{acao1b}
\end{equation}
We notice in passing that, due to the inability of general relativity to deal with the lack
of universality, this action cannot be reduced to the general relativistic form
(\ref{graction}). Variation of (\ref{acao1b}) yields \cite{wep}
\begin{equation}
\left( \partial_\mu x^a +
\frac{m_g}{m_i} \; B^a{}_\mu \right) \frac{d u_a}{d s} =
\frac{m_g}{m_i} \; \Tw^a{}_{\mu \rho} \; u_a \, u^\rho.
\label{eqmot22}
\end{equation}
This is the force equation governing the motion of the particle, in which the teleparallel
field strength $\Tw^a{}_{\mu \rho}$ plays the role of gravitational force. Similarly to
the electromagnetic Lorentz force, which depends on the relation $q/m_i$, with $q$ the
electric charge, the gravitational force depends explicitly on the relation ${m_g}/{m_i}$
of the particle. In the Newtonian limit, this force equation reduces to the original
Newton's law
\begin{equation}
m_i \frac{d^2 \vec{x}}{d t^2} = - m_g \vec{\nabla} \phi,
\end{equation}
with $\phi = c^2 B_{00}$ the gravitational potential \cite{global}. It is important to
observe that this limit is possible only because both teleparallel and Newtonian theories
are able to manage with the absence of universality, which is not the case of general
relativity. For this reason we can say that Newton's theory follows much more naturally
from teleparallel gravity than from general relativity.

The crucial point is to observe that, although the equation of motion depends explicitly
on the relation $m_i/m_g$ of the particle, neither $B^a{}_\mu$ nor $\Tw^a{}_{\rho 
\mu}$ depends on this relation. This means essentially that the teleparallel field
equation  (\ref{tfe1}) can be consistently solved for the gravitational potential
$B^a{}_\mu$, which can then be used to write down the equation of motion (\ref{eqmot22}),
independently of the validity or not of the weak equivalence principle. Even in the 
absence of universality, therefore, teleparallel gravity is able to consistently describe
the gravitational interaction \cite{wep}.

Let us now see what happens in the context of general relativity. By using the identity
(\ref{tuukuu}), the force equation (\ref{eqmot22}) can be rewritten in the form
\begin{equation}
\frac{d u_\mu}{ds} - \Gammabol^\lambda{}_{\mu \rho} \, u_\lambda \, u^\rho =
\left(\frac{m_g - m_i}{m_g} \right) \partial_\mu x^a \, \frac{d u_a}{d s},
\label{eqmot6}
\end{equation}
where use has been made of the relation (\ref{rela}). Notice that the  violation of
the weak equivalence principle produces a deviation from the geodesic motion, which is
proportional to the difference between the gravitational and inertial
masses.\footnote{Notice that, due to the assumed non-universality of free fall, it is not
possible to find a local coordinate system in which the equation (\ref{eqmot6})
reduces to the equation of motion of a free particle. This is a consequence of the fact
that a violation of the weak equivalence principle precludes the existence of a strong
version of the principle.} Of course, when $m_g = m_i$, the equation of motion
(\ref{eqmot6}) reduces to the geodesic equation of general relativity. However, in the
absence of the weak equivalence principle, it is not a geodesic equation, which means that
it does not comply with the {\em geometric} description of general relativity, according
to which the trajectories of all (spinless) particles must be given by {\em geodesics}.
Furthermore, in the context of general relativity, there is no an action integral from where
(\ref{eqmot6}) can be obtained through a variational principle.

In order to comply with the foundations of general relativity, it is necessary to
incorporate the particle properties into the geometry. This can be achieved by assuming,
instead of the tetrad (\ref{tetrada})  of teleparallel gravity, the new tetrad
\begin{equation}
\bar{h}^a{}_\mu = \partial_\mu x^a + \frac{m_g}{m_i} \; B^a{}_\mu,
\label{tetrada2}
\end{equation}
which takes into account the relation $m_g/m_i$ of the particle under consideration.
This tetrad defines a new spacetime metric tensor
\begin{equation}
\bar{g}_{\mu \nu} = \eta_{a b} \; \bar{h}^a{}_\mu \; \bar{h}^b{}_\nu,
\label{gmn2}
\end{equation}
in terms of which the corresponding spacetime invariant interval is
\begin{equation}
d\bar{s}^2 = \bar{g}_{\mu \nu} \, dx^\mu dx^\nu.
\end{equation}
By noticing that in this case the relation between the gravitational field  strength and
torsion turns out to be
\begin{equation}
\frac{m_g}{m_i} \; T^a{}_{\mu \rho} = \bar{h}^a{}_\lambda \, 
\bar{T}^\lambda{}_{\mu \rho},
\label{fstor2}
\end{equation}
it is an easy task to verify that, for a fixed relation $m_g/m_i$, the equation of  motion
(\ref{eqmot2}) is equivalent to the true geodesic equation
\begin{equation}
\frac{d \bar{u}_\mu}{d\bar{s}} - {\bar{\Gamma}}{}^\lambda{}_{\mu \rho}
\, \bar{u}_\lambda \, \bar{u}^\rho = 0,
\label{eqmot7}
\end{equation}
where $\bar{u}_\mu \equiv d x_\mu/d \bar{s} = \bar{h}^a{}_\mu u_a$, and
$\bar{\Gamma}{}^{\rho}{}_{\mu\nu}$ is the Christoffel connection of the metric
$\bar{g}_{\mu \nu}$. Notice that this equation can also be obtained from the  action
integral
\begin{equation}
\bar{S} = - m_i \ c \int_a^b d\bar{s},
\end{equation}
which is the usual form of the action in the context of general relativity.

However, the price for imposing a geodesic equation of motion to describe a non-universal
interaction is that the gravitational theory becomes inconsistent. In fact, the  solution
of the corresponding Einstein's field equation
\begin{equation}
\bar{R}_{\mu \nu} - \frac{1}{2} \, \bar{g}_{\mu \nu} \bar{R} =
\frac{8 \pi G}{c^4} \, \bar{\mathcal T}_{\mu \nu},
\label{e2}
\end{equation}
with $\bar{\mathcal T}_{\mu \nu} = \delta {\mathcal L}_m/\delta \bar{g}^{\mu \nu}$, would
in this case depend on the relation $m_g/m_i$ of the test particle, which  renders the
theory inconsistent in the sense that test particles with different relations 
$m_g/m_i$ would require connections with different curvatures to keep all equations of 
motion given by geodesics. Of course, the gravitational field cannot depend on  any test
particle properties. We can then conclude that, in the absence of the weak equivalence
principle, the geometric description of general relativity breaks down. Since the gauge
potential $B^a{}_\mu$ can always be obtained independently of any property of the test
particle, teleparallel gravity remains as a consistent theory in the lack of universality.

%%%%%%%%%%%
\section{Conclusions}
%%%%%%%%%%%

One of the fundamental problems of gravitation is the conceptual conflict of Einstein's
general relativity with quantum mechanics. Technically, it usually shows up as the
impossibility of obtaining a renormalizable quantum theory for gravitation. However, there
are fundamental reasons behind such inconsistency, essentially  related to the very
principles on which these theories are based. General relativity, as is well known, is
based on the equivalence principle, whose strong version establishes the {\it local}
equivalence between gravitation and inertia. Quantum mechanics, on the other hand, is
fundamentally based on the uncertainty principle, which is a {\it nonlocal} principle. On
this fundamental difference lies one of the roots of the difficulty in reconciling these
two theories \cite{chiao}.

Now, although equivalent to general relativity, the gauge approach of teleparallel gravity
is able to describe gravitation in a consistent way without resorting to the equivalence
principle. The crucial point is the different character of the fundamental field of each
theory: whereas in general relativity it is a tetrad field $h^a{}_\mu$ (or equivalently, a
metric tensor $g_{\mu \nu}$), in teleparallel gravity it is a gauge potential $B^a{}_\mu$,
the nontrivial part of the tetrad field:
\begin{equation}
h^a{}_\mu = \partial_\mu x^a + B^a{}_\mu.
\label{ft1}
\end{equation}
This apparently small difference has deep consequences. In fact, any gravitational theory
whose fundamental field is a tetrad (or a metric) is necessarily a {\em  geometrical theory}.
On the other hand, a theory whose fundamental field is a gauge potential has not the same
geometrical character. As a gauge theory it can, similarly to Maxwell's theory,  be
formulated independently of any equivalence principle. To understand this point, let us
consider a particle whose gravitational mass $m_g$ does not coincide with its inertial mass
$m_i$. Of course, both the weak and the strong equivalence principles are no longer valid. In
this case, as we have seen, a geometrical theory for gravitation would require the
introduction of a new tetrad field, given by \cite{wep}
\begin{equation}
\bar{h}^a{}_\mu = \partial_\mu x^a + \frac{m_g}{m_i} \, B^a{}_\mu.
\label{ft2}
\end{equation}
Since the relation ${m_g}/{m_i}$ of the test particle appears ``inside'' the  tetrad
definition, any theory in which $\bar{h}^a{}_\mu$ is the fundamental field will be
inconsistent in the sense that particles with different relations ${m_g}/{m_i}$ would
require connections with different curvatures to keep a geometric description of
gravitation, in which all trajectories would be given by geodesics. On the other
hand, we  see from the tetrad (\ref{ft2}) that the relation ${m_g}/{m_i}$ appears
``outside'' the gauge potential $B^a{}_\mu$. This means essentially that, in this case, the
gravitational field equations (\ref{tfe1}-\ref{emt1bis}) can be consistently solved for
$B^a{}_\mu$ independently of any test-particle property. This is the fundamental reason for
teleparallel gravity to remain as a viable theory for gravitation, even in the absence of
universality. We can then conclude that, similarly to what happens in Maxwell's theory,
which is also a gauge theory, teleparallel gravity does not require the existence of an
equivalence principle to describe the gravitational interaction.\footnote{We observe that,
from the point of view of teleparallel gravity, the question of the existence or not of a
quantum version of the weak equivalence principle becomes irrelevant.} The replacement of
general relativity by teleparallel gravity, therefore, may lead to a conceptual
reconciliation of gravitation with quantum mechanics. Accordingly, the quantization of the
gravitational field may also appear more consistent if considered in the teleparallel
picture.  This is, of course, an open question yet to be explored.

%%%%%%%%%%%%%%
\section*{Acknowledgments}
The authors would like to thank FAPESP, CNPq and CAPES for financial support.

%%%%%%%%%%%%%%

\end{document}